\documentclass[aps,prd,showpacs]{revtex4}
\usepackage{graphicx}
\usepackage{color}
\usepackage{latexsym}
\usepackage{amsmath}
\usepackage{amsfonts}
\usepackage{amssymb}
\usepackage{verbatim}

\newcommand{\be}{\begin{equation}}
\newcommand{\ee}{\end{equation}}
\newcommand{\bea}{\begin{eqnarray}}
\newcommand{\eea}{\end{eqnarray}}

\newcommand{\ben}{\begin{eqnarray}}
\newcommand{\een}{\end{eqnarray}}

\begin{document}

\title{Domain wall description of superconductivity}
\author{ F. A. Brito$^{1}$, M. L. F. Freire$^{2}$, J. C. Mota-Silva $^{1,3}$}
\affiliation
{$^{1}$Departamento de F\'\i sica,
Universidade Federal de Campina Grande, Caixa Postal 10071,
58109-970  Campina Grande, Para\'\i ba, Brazil\\
$^{2}$Departamento de F\'\i sica, Universidade Estadual da Para\'\i ba, 58.109-753 Campina Grande, PB, Brazil\\
$^{3}$Departamento de F\'\i sica, Universidade Federal da Para\'\i ba, Caixa Postal 5008, 58051-970 Jo\~ao Pessoa, Para\'\i ba, Brazil}


\begin{abstract}
In the present work we shall address the issue of electrical conductivity in superconductors in the perspective
of superconducting domain wall solutions in the realm of field theory. 
We take our set up made out of a dynamical complex scalar field coupled to gauge field to be responsible
for superconductivity and an extra scalar real field that plays the role of superconducting domain walls. 
The temperature of the system is interpreted through the fact that the soliton following accelerating orbits is a Rindler observer experiencing a thermal bath.
\end{abstract}

\maketitle


\section{Introduction}


In this paper we present an alternative description of superconductivity in field theory. We take advantage of the rich and well stablished features involving classical soliton solutions, mainly those formulated in multi-scalar fields \cite{witten, vilenkin, mackenzie,bazeia}. The reason for using such solutions in our purpose is twofold. Firstly, because they can develop internal structures such as condensates which are fundamental for studying superconductivity. Alternatively, this opens the possibility of studying {\it superconducting solitons}. The first example of such objects were the superconducting strings \cite{witten,vilenkin} ---  other developments in domain walls with internal structures were also considered in \cite{mackenzie,bazeia}. Secondly, because these solutions can follow non-trivial orbits in the field space \cite{bazeia}. They mostly force the solitons to move into accelerated trajectories. As such, we can identify these solitons as non-relativistic Rindler observers experiencing a thermal bath. As we shall see, this will be fundamental to introduce temperature in the system in a very natural way and identify several important quantities such as the condensate and resistivity as a function of the temperature. We believe this alternative can open a new window for investigating superconductivity in field theory through superconducting solitons since there exist many types of soliton solutions in many well established field theories such a way superconducting solitons can also be identified. This new perspective may complement and shed some new light on earlier studies of High-$T_c$ superconductivity in field theory focused on `particle excitations' only \cite{pwa}. 

In the present study, we take our set up made out of a dynamical complex scalar field coupled to the abelian gauge field to be responsible for superconductivity and an extra real scalar field that plays the role of superconducting domain walls.
They are domain wall backgrounds that develop a condensate in their core. 
Furthermore, the domain wall model is a good approximation for superconductors presenting, for instance, a layer-type perovskite-like structure \cite{bm1986, tinkham}.
The quantum field theory can explain some effects of superconductivity. The results can be obtained with an appropriate {classical regime of a} quantum field theory inspired by the Ginzburg-Landau (GL) theory \cite{GL}. Though initially proposed as a phenomenological theory, the GL  theory can be shown to be a limiting case of a microscopic theory \cite{gorkov} such as the  Bardeen-Cooper-Schrieffer (BCS) \cite{BCS}  theory of superconductivity. The domain of validity of the GL theory is shown to be restricted to temperatures sufficiently near the critical temperature and to spatially slow varying fields \cite{tinkham}. 

The paper is organized as follows. In Sec.~\ref{supII} we address the issue of type II domain wall solutions in field theory coupled to gauge field. We consider such solutions as backgrounds fields to solve a Schroedinger-like equation for the electromagnetic field. In Sec.~\ref{cond-Temp} we compute the condensate at finite temperature. The temperature of the system is interpreted through the fact that the soliton following accelerating orbits is a Rindler observer experiencing a thermal bath. In Sec.~\ref{conduct-ac} we calculate the optical conductivity in terms of the frequency and temperature. In the limit of very low frequencies and temperatures the superconductor develops infinite DC conductivity, as expected. Furthermore, in this regime we can easily read off the binding energy $\Delta$ of a Cooper pair from the real part of the optical conductivity. At some temperature around the critical temperature the AC resistivity develops an effect commonly seen in High$-T_c$ superconductors \cite{bm1986}.  Finally, in Sec.~\ref{conclu} we make our final comments.

\section{Superconducting Type II domain walls solution}
\label{supII}

\begin{figure}[h!]
        \includegraphics[{angle=90,height=5.8cm,angle=270,width=6.8cm}]{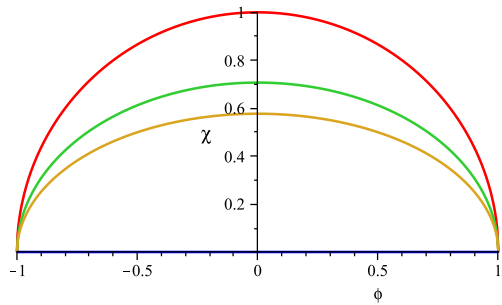}  \qquad\qquad 
  \caption{Elliptic orbits followed by the type II kink. The temperature becomes high from the top ($T=0$) to bottom ($T=T_c$) orbits.}
  \label{fig-orbits}
\end{figure}

To obtain superconducting domain walls is needed a complex scalar field with charge $q$ that must couple to a real scalar field that produces the domain walls. By introducing a coupling between the complex scalar field and the electromagnetic field, the former develops a {\it condensate} inside the domain wall which becomes superconducting and develops almost all the properties of a superconducting material. The superconducting domain wall developing a condensate can be generated by the following Lagrangian with the $Z_2\times U(1)$ symmetry:
\ben
\mathcal{L} &=& \frac{1}{2}\partial_{\mu}\phi\partial^{\mu}\phi + (\partial^{\mu}\chi + iqA^{\mu}\chi)(\partial_{\mu}\chi^{*} - iqA_{\mu}\chi^{*})\nonumber\\
& -& V(\phi,\chi, \chi^{*}) - \frac{1}{4}F_{\mu \nu}F^{{\mu \nu}},
\label{london 2}
\een
where $\mu,\nu=0,1,2, ..., d$ are bulk indices for arbitrary $(d-2)$-dimensional domain walls. We shall focus on two-dimensional domain walls with bulk indices  running as $\mu,\nu=t,x,y,r$.
The potential $V(\phi, \chi, \chi^{*})$ is chosen appropriately so that the domain wall becomes a superconductor
\ben
V(\phi, \chi,\chi^{*}) = \frac{1}{2}\lambda^{2}(\phi^{2}-a^{2})^{2}+\lambda\mu(\phi^{2} \nonumber\\
- a^{2})\left|\chi\right|^{2}+\frac{1}{2}\mu^{2}\left|\chi\right|^{4} + \mu^{2}\phi^{2}\left|\chi\right|^{2}.
\label{london 7}
\een
The real scalar field $\phi$ develops $Z_2$ symmetry and is responsible to form the domain wall, whereas the charged scalar field $\chi$ develops a condensate inside the domain wall and is also responsible to produce type II domain walls. 
The equations of motion for the real and complex scalar fields coupled to electromagnetic field are given by
\ben
\label{eom-phi}
&&\square\phi+\frac{\partial V}{\partial\phi}=0,\\
\label{eom-chi}
&&\square\chi+\frac{\partial V}{\partial\chi^{*}}-2iqA_\mu\partial^\mu\chi-q^2A_\mu A^\mu\chi=0, \qquad c.c.,\\
\label{EM}
&&\square A_{\mu}+iq(\chi^{*}\partial_\mu\chi-\chi\partial_\mu\chi^{*})+2q^2A_\mu|\chi|^2=0.
\een
For $A_\mu=0$ the scalar real sector produces domain wall solutions whose kink profiles are the following well-known BPS {\it static} solutions obtained in terms of first order formalism and a specific superpotential \cite{bazeia}.
The type I solution
\ben\label{typeI}
&&\phi(r)=-a\tanh{(\lambda a r)},\nonumber\\
&&\chi=0,
\een
and the type II solution
\ben\label{typeII}
&&\phi(r)=-a\tanh{(2\mu a r)},\nonumber\\
&&\chi(r)=\pm a\,\sqrt{\frac{\lambda}{\mu}-2}\,\mbox{sech}\,{(2\mu ar)},
\een
where $r$ is the spatial coordinate transverse to the domain walls. These solutions correspond to straight and elliptic orbit, respectively \cite{bazeia} --- See Fig.~\ref{fig-orbits}. 
They have the same Bogomol'nyi energy. Note that for sufficiently large $\lambda/\mu$ the elliptic orbit 
\ben\phi^2+{\Big(\frac{\lambda}{\mu}-2\Big)}^{-1}{\chi^2}=a^2,\een
 passes through the `supersymmetric vacuum' $\phi=0$ and $\chi=\pm\sqrt{\lambda/\mu}$ --- global minima of the scalar potential written in terms of first derivatives of superpotential \cite{bazeia}. As we shall discuss later, this will correspond to zero temperature, which agrees with the formation of a condensate inside the superconducting domain wall.

Now considering these solutions  as background fields we solve the equation of motion for electromagnetic field (\ref{EM}) on these backgrounds. By introducing $A_\mu(t,r)=A_\mu(r)e^{-i\omega t}$ and $\chi(t,r)=\chi(r)e^{-i\theta t}$ we obtain the Schroedinger-like equation for $A_x$ (or  $A_y$) as follows
\ben\label{flut-a}
-A^{\prime\prime}_x+\frac14\ell^2\mbox{sech}^2\,{(\alpha r)}\,A_x=\omega^2A_x,
\een
where $\ell=2\sqrt{2}qa \sqrt{\frac{\lambda}{\mu}-2}$ and $\alpha=2\mu a$. This is a well-known Schroedinger problem with a \mbox{sech}-type  barrier potential whose solution is given by
\ben\label{A_sol}
A_x(\omega,\alpha,\ell,r)= \Big( {\rm sech}\left( \alpha\,r \right)\Big)^{-\frac{i\omega}{\alpha}}{_2F_1}\left[{\it a1},{\it a2};\,{\it a3};\,\frac12(1-\,\tanh \left( \alpha\,r \right)) \right],
\een
where $_2F_1$ is a hypergeometric function with the parameters defined as
\ben\label{a-parameters}
&&{\it a1}=\frac12\frac{-2i\omega+\alpha+\sqrt{-\ell^2+\alpha^2}}{\alpha} , \nonumber \\
&&{\it a2}=-\frac12\frac{2i\omega-\alpha+\sqrt{-\ell^2+\alpha^2}}{\alpha}, \\
&&{\it a3}= -\frac{i\omega-\alpha}{\alpha}.  \nonumber
\een

\section{The condensate at finite temperature}
\label{cond-Temp}
{The usual manner to introduce finite temperature in quantum field theory is through path integral with Euclidean time whose period is related with the temperature of the system. In this sense the path integral turns out  to be the partition function from which one can obtain Bose-Einstein and Fermi-Dirac statistics for bosonic and fermionic fields, respectively. Furthermore, in curved spacetime or equivalently for accelerated observers such statistics tell us that the temperature is related to surface gravity or acceleration --- see \cite{unruh}  for further details. In our system although the superconducting domain wall is living in a flat spacetime its related soliton (i.e., its BPS kink profile (\ref{typeII})) is accelerated in the field space as a non-relativistic Rindler observer --- see below. As we have previously mentioned we shall take advantage of this fact to identify the temperature of the system as in the following.
We shall first attempt to convince the reader that the parameter $\alpha$ can be indeed related to the temperature of the system. }
The orbits of Fig.~\ref{fig-orbits} followed by the BPS solutions into the $(\phi,\chi)$-plane  make the type II kink solutions to experiment {constant} {\it accelerations}. { One can think of such soliton solutions following accelerating trajectories (or orbits). We can also see these trajectories in a `Lorentzian' signature in $(\phi,\chi)$-plane by making the change $\phi\to i\phi$ --- see Fig.~\ref{hyper} --- and $r\to i\tau$ in the BPS solutions (\ref{typeII}). 

The soliton in this sense is a Rindler observer experiencing a thermal bath of bosonic and fermionic modes
 that are distributed, respectively, according to Bose-Einstein and Fermi-Dirac statistics  (this is the Unruh effect \cite{unruh})
\ben
<n_{\Omega}>=\frac{1}{\exp{\left(\frac{2\pi\Omega}{a}\right)}\mp 1},
\een
where the constant acceleration $a$ is promptly identified with the Unruh temperature $T=\frac{a}{2\pi}$ and $\Omega$ are the frequencies of the {\it thermal radiation}.}
{
\begin{figure}[h!]
        \includegraphics[{angle=90,height=5.8cm,angle=270,width=6.8cm}]{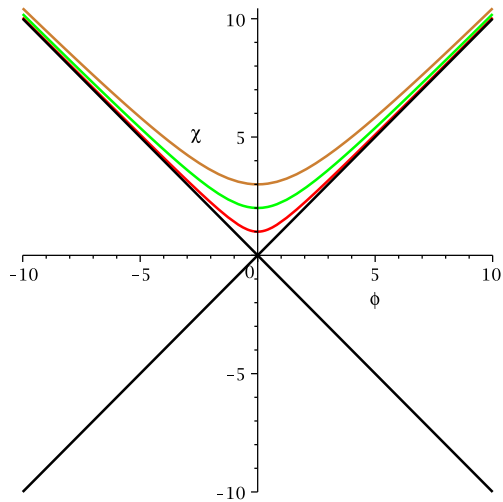}  \qquad\qquad  
  \caption{ Hyperbolic worldlines (orbits) followed by the type II solitons in `Lorentzian' signature of the ($\phi,\chi$)-space.}
  \label{hyper}
\end{figure}
In the following we show that the solution (\ref{typeII}) indeed describes an accelerated trajectory. Firstly, we redefine the fields $\phi, \chi$ in terms of coordinates of space and time as follows $\phi=\alpha a\, t(\tau)$ and $\chi=\alpha a\left({\frac{\lambda}{\mu}-2}\right)^{1/2} z(\tau)$ such that
\ben
\label{coords-t-x}
&&t(\tau)=\frac{1}{\alpha}\tanh{\alpha \tau},\nonumber\\
&&z(\tau)=\frac{1}{\alpha}{\rm\, sech\,}{\alpha \tau},
\een  
where $\tau\equiv x$ is identified with (Euclidean) proper time. Now using the definition of the acceleration $a^\mu=\frac{d^2 x(\tau)^\mu}{d\tau^2}$ we find that 
\ben
a\equiv\sqrt{a_\mu a^\mu}&=&\alpha-\frac12\alpha(\alpha\tau)^2+...\nonumber\\
&\simeq&\alpha.
\een
In the last step we have assumed the regime of very slow velocities $v=a\tau\ll1$ (non-relativistic Rindler observer). Thus, using the Unruh temperature we find $\alpha\simeq 2\pi T$ that we shall simply assume throughout the paper the useful correspondence
\ben\label{alpha_T}
\alpha\equiv T.
\een

 }
On the other hand, let us now compute the analog of acceleration in the $(\phi,\chi)$-plane. Actually we will be indeed calculating the inverse of acceleration since the coordinates $(\phi,\chi)$ have dimension of energy instead of length as in the usual sense. We can now define the acceleration of the system as 
\ben\label{acelera-1}
a(r_0)^{-1}=\left.\frac{d^2\chi}{d\phi^2}\right |_{r_0},
\een
where $r_0$ is some point on the bulk.  
 The r.h.s. of  Eq.~(\ref{acelera-1}) can be written in terms of the superpotential \cite{bazeia} and { type II kink solutions} in the form
\ben\label{acelera-2}
\left.\frac{d^2\chi}{d\phi^2}\right |_{r_0}=\left.\frac{d}{d\phi}\left(\frac{d\chi}{d\phi}\right)\right |_{r_0}=\left.\frac{d}{d\phi}\left(\frac{W_\chi}{W_\phi}\right)\right |_{r_0}=\left.\frac{W_{\phi\chi}}{W_\phi}-\frac{W_{\phi\phi}W_\chi}{W_\phi^2}\right |_{r_0}=-\frac{\sqrt{\frac{\lambda}{\mu}-2}}{a\,{\rm sech\,}{(2\mu a  r_0)}}\left( 1+\frac{\lambda}{\mu}\frac{\tanh^2{(2\mu a r_0)}}{\,{\rm sech\,}^2{(2\mu a r_0)}}\right),
\een 
where the minus sign simply reflects the concavity of the orbits. The temperature can be now defined in the general form 
\ben
\beta=\left| \frac{d^2\chi}{d\phi^2}\right |_{r_0},\qquad \beta=\frac{1}{T}.
\een
In order to identify the temperature in terms of the parameters of the theory let us investigate the regime near the core of the domain wall (i.e., in the supersymmetric vacuum $\phi=0$ and $\chi=\pm\sqrt{\lambda/\mu}$), that is $r_0\approx0$ and $\lambda/\mu\gg1$. This means that the temperature is now defined through the following equation 
\ben\label{temperature}
\frac{1}{T}=\frac{\sqrt{\frac{\lambda}{\mu}-2}}{a}\approx \frac{\sqrt{\frac{\lambda}{\mu}}}{a},
\een
{The equations (\ref{alpha_T}) and (\ref{temperature}) together with the fact that $\lambda a\equiv T_c$ (that by construction we assume to be the regime of critical temperature, since for $T<T_c$ we have shown that $2\mu a=T$), 
are easily satisfied by the the dimensionless ($\lambda, \mu$) and dimensionful ($a$) parameters, up to numeric irrelevant pre-factors, as follows}
\ben
\label{condsat-T}
\lambda\sim \frac{T_c^{1/2}}{T^{1/2}}, \qquad \mu\sim \frac12\frac{T^{1/2}}{T_c^{1/2}}, \qquad a\sim T_c^{1/2}T^{1/2}.
\een

\begin{center}
\begin{figure}[h!]
        \includegraphics[{angle=90,height=6.2cm,angle=270,width=7.2cm}]{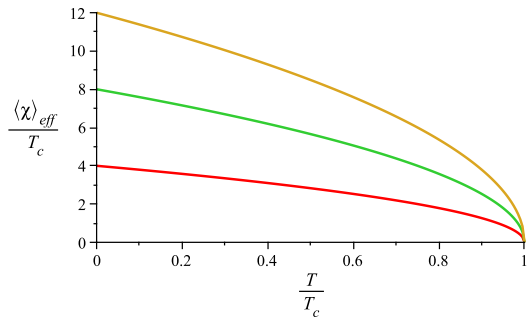}  
  \caption{
  The effective condensate as a function of temperature for charges $q=1, 2$, and 3 from bottom to top.}
  \label{fig1}
\end{figure}
\end{center}

Let us now consider a region out of the core,  that is $2\mu a r_0\ll1$ and $\lambda/\mu\gg1$. Now substituting these assumptions into (\ref{acelera-2}) we obtain the temperature
\ben\label{temperature-BCS}
T=\frac{a(\lambda/\mu)^{-3/2}}{(2\mu a r_0)^2}.
\een
Now the temperature has a dependence with four parameters. However, we shall keep { using the correspondence} $\alpha=2\mu a\equiv T$ and $\lambda a\equiv T_c$ as in the previous analysis. Thus, substituting again the parameters (\ref{condsat-T}) into the temperature (\ref{temperature-BCS}) we readily find a relation between the critical temperature $T_c$ and the scale $r_0$ given by
\ben\label{temperature-Tc}
T=\frac{1}{2^{3/2}T_cr_0^2}.
\een
Since by definition the temperature $T$ does not depend on $r_0$ then 
\ben
T_cr_0^2=const.
\een
{We are still tempted to write this formula in terms of atomic mass $A$ in the bulk since the lattice parameter can be well approximated by  $r_0\simeq A^{1/3}\,fm$. Thus, we arrive to the {\it isotopic mass formula} \cite{tinkham}
\ben
T_c A^{2/3}=const.
\een
}
The condensate can be easily isolated by power expanding the scalar solution $\chi(r)$ in Eq.~(\ref{typeII}) around the core of the type II domain walls at $r\approx0$ as
\ben\label{condsat0}
\chi(r)=m-\frac{1}{2}m\alpha^2r^2+...,
\een
with the condensate given by $<\chi>\simeq m$.  We have written the original type II solution as $\phi(r)=-a\tanh{\alpha r}$, $\chi(r)=m\,\mbox{sech}{\,\alpha r}$. 
To find the explicit dependence of the condensate with the temperature we recall the original form of the parameters $m,\alpha$, that is
\ben
\label{condsat}
m=a\sqrt{\frac{\lambda}{\mu}-2},\qquad \alpha=2\mu a.
\een
Now substituting (\ref{condsat-T}) into (\ref{condsat}) allows us to obtain
\ben
\label{condsat-TT}
m=\sqrt{2}T_c\sqrt{1-\frac{T}{T_c}},
\een
which implies that the condensate has precisely the desired form $<\chi>\simeq m=\sqrt{2}T_c\sqrt{1-{T}/{T_c}}$. We also note that from the equation of motion of electromagnetic field $A_x$ the {\it effective condensate} `seen' by the electromagnetic field is given in terms of the charge $q$, i.e., $<\chi>_{eff}\simeq\ell=2\sqrt{2}\,q\,m$ or simply $<\chi>_{eff}\simeq4\,q\,T_c\sqrt{1-{T}/{T_c}}$, { where $\ell$ is defined just below Eq.~(\ref{flut-a})} ---  See Fig.~\ref{fig1} for the explicit behavior of the condensate $<\chi>_{eff}$ with the temperature.   

\section{Conductivity}
\label{conduct-ac}

As well-known, from the Ohm's law we can readily obtain the conductivity along a direction, say $x$-direction along the domain walls, in the form
\ben
\sigma_x (x,y) = \frac{J_{x}}{E_{x}} = \frac{A'_x(0)}{i\omega A_x(0)},
\label{ohm 2 - new}
\een
where in the last step we used $E_x=-\partial_t A_x =i\omega A_x$, with $A_x(t,r)=A_{x}(r) e^{-i\omega t}$ and defined the current as $J_x=A'_x(0)$.
This can be justified by using the boundary conditions for the electromagnetic field on an interface at $r=0$ (or in other position as we shall consider later) that corresponds to a plane along the superconducting domain wall.  More specifically, the boundary conditions for the magnetic field at an interface is
\ben\label{bdary}
\hat{n}\times \vec{B}=\vec{J}, \qquad\mbox{at $r=0$}
\een 
where $\hat{n}$ is a normal vector to the surface of the domain wall and $\vec{J}$ is a surface current. For $\hat{n}=(0,0,1)$ and $\vec{A}=(A_x, A_y, 0)$ the boundary condition (\ref{bdary}) simply becomes
\ben
-\partial_rA_x(r)=J_x,  \qquad\mbox{at $r=0$}
\een
which is nothing but our anticipated assumption $J_x=A'_x(0)$.

\begin{figure}[h!]
        \includegraphics[{angle=90,height=5.8cm,angle=270,width=6.8cm}]{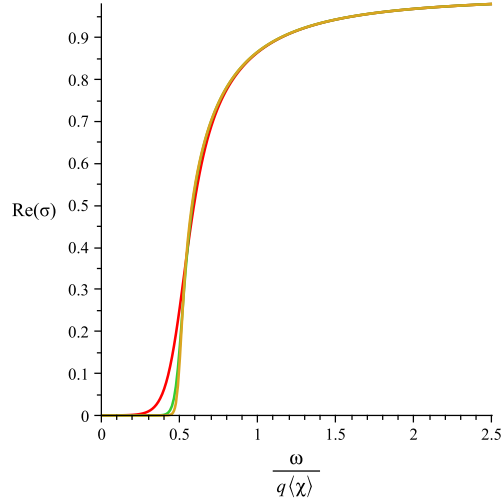}  \qquad\qquad 
  \caption{
  The real part of the conductivity as a function of the frequency normalized by the effective condensate. 
  We use the charges $q=8, 20$, and 32 from top to bottom;  $\delta = 0.01$ and $\ell\simeq <\chi>_{eff}=4$.}
  \label{fig2}
\end{figure}

Now using the solution (\ref{A_sol}) for the electromagnetic field 
around a generic point $r\approx \delta$  (sames as $r_0$) we are able to write the explicit form of the conductivity $\sigma_x=\sigma_y\equiv\sigma$ as follows
\ben\label{sigma_final}
\sigma(\omega,\alpha,\ell,\delta)={\frac {\frac18\,i \left( 4\,{\omega}^{2}+4\,i\omega\,\alpha-{\ell}^{2}
 \right) 
{_2F_1\left[{\it b1},{\it b2};\,{\it b3};\,\frac12(1-\,\tanh \left( \alpha\,\delta \right)) \right]}
 {\rm sech}^{2}\left( \alpha\,\delta \right)}{\omega
\, \left( i\omega-\alpha \right) 
{_2F_1\left[{\it a1},{\it a2};\,{\it a3};\,\frac12(1-\,\tanh \left( \alpha\,\delta \right))\right ]}
}}+\tanh \left( \alpha\,\delta \right),
\een
with the parameters ${\it b1}, {\it b2}$ and ${\it b3}$ defined as 
\ben\label{sigma_final2}
&&{\it b1}=-\frac12\frac{2i\omega-3\alpha+\sqrt{-\ell^2+\alpha^2}}{\alpha},\nonumber \\
&&{\it b2}=\frac12\frac{-2i\omega+3\alpha+\sqrt{-\ell^2+\alpha^2}}{\alpha},\\
&&{\it b3}= -\frac{i\omega-2\alpha}{\alpha} \nonumber.
\een
Recall we have previously defined the temperature  $\alpha\equiv T$ and the condensate $m\simeq<\chi>$. We now consider the conductivity normalized by the `effective condensate' $\ell\to q\ell$, such that we define $\alpha=q^{-1} q\, \ell$ and $\omega=\omega_r q\, \ell$ into $\sigma$. We can still write $\frac{\alpha}{q <\chi>_{eff}}=q^{-1}$ and  $\frac{\omega}{q <\chi>_{eff}}=\omega_r$ (reduced frequency). Finally we substitute all over this into (\ref{sigma_final})-(\ref{sigma_final2}).
The results have shown that for $\delta\approx0$ the optical conductivity --- See Fig.~\ref{fig2} --- is essentially the same as the one computed at $r=0$, i.e., at the core of the domain wall.  On the other hand, as we shall see later, the conductivity (or AC resistivity) as a function of the temperature is more sensitive to the values of $\delta$. In the following we focus on further characteristics of the optical conductivity  by simply  assuming $\alpha\delta=0$.

As $\omega\to0$ and $T\to 0$ the conductivity (\ref{sigma_final}) approaches a delta function  $\delta(\omega)$. This is because for $T\to0$, we have $\ell\sim T_c$. 
Thus, in this limit, $\alpha^2\ll \ell^2$ and $\omega^2\ll \ell^2$  implies that the real and imaginary parts of the conductivity, up to a prefactor  $\sim2\alpha/\ell$ from the hypergeometric functions, can be written as 
\ben\label{distrib}
{\rm Re}\,\sigma(\omega)\propto\frac{(\ell/\alpha)^2}{(\omega/\alpha)^2+1}\to\delta(\omega), \qquad {\rm Im}\,\sigma(\omega)\propto\frac{(\ell/\alpha)^2}{(\omega/\alpha)^3+\omega/\alpha} \to\frac{\ell^2}{\alpha}\frac{1}{\omega}.
\een
Note that the distribution in real part of (\ref{distrib}) tends to a delta function, whereas the imaginary part presents a pole. This in accord with the Kramers-Kronig relations and 
Drude model of conductor in the limit of the relaxation time due to scattering $\tau\to\infty$ (superconductor).  We conclude from such discussion that in the limit $\omega\to0$ at $T\to0$ our model presents an infinite DC conductivity as expected for a superconductor.

Now let us consider the conductivity as a function of the temperature. Repeating the previous analysis for $\alpha\delta\to\infty$ the argument in the hypergeometric functions goes to zero as $e^{-2\alpha \delta}$. In this regime the ratio of the hypergeometric functions in the conductivity formula can be well approximated by a series of a few terms. By keeping only leading and next-to-leading terms we find 
 \ben\label{distrib2}
{\rm Re}\,\sigma(\omega,\alpha)\propto\delta(\omega)\left(1-\frac18\frac{\ell^2}{\alpha^2}e^{-2\alpha \delta}+...\right)\simeq \delta(\omega)e^{-\frac18\left(\frac{\Delta}{\alpha}\right)^2},\een
where 
\ben\label{Delta-liga}
\Delta={\ell}e^{-\alpha \delta},
\een
precisely defines the binding energy of a Cooper pair as long as we identify $\ell=2\omega_D$ as the Debye temperature and $\delta\alpha=1/VN_F$, being $V>0$ the binding potential and $N_F$ the density of orbitals with Fermi's energy. Note that the limit $\delta\alpha\to\infty$ corresponds to $VN_F\to0$ that is the limit of weak coupling which is in accord with the BCS theory. { On the other hand, the limit $\delta\alpha\to0$ corresponds to $VN_F\to\infty$ that is the limit of strong coupling.}

In order to go further into our analysis we extend the power expansion of the condensate in (\ref{condsat0}) to the expansion around a plane $r\approx\delta$, parallel to the domain wall near its center given by 
\ben\label{condsat01}
\chi(r)=m\,{\rm sech }{(\alpha\delta)}-m\,{\rm sech }{(\alpha\delta)}\tanh{(\alpha\delta)}\alpha(r-\delta)+...,
\een
which allows us to redefine the usual condensate previously studied in the form $<\chi>\simeq m \,{\rm sech}\,{(\alpha\delta)}=\sqrt{2}T_c\sqrt{1-{T}/{T_c}}\,{\rm sech}\,{(\alpha\delta)}$ and the effective condensate as  $<\chi>_{eff}\simeq\ell \,{\rm sech}\,{(\alpha\delta)}=2\sqrt{2}\,q\,m \,{\rm sech}\,{(\alpha\delta)}$ or simply $<\chi>_{eff}\simeq4qT_c\sqrt{1-{T}/{T_c}}\,{\rm sech}\,{(\alpha\delta)}$. In the regime $\alpha\delta\to\infty$, particularly,  we write the effective condensate given by
\ben\label{cond-liga}
<\chi>_{eff}\simeq2\ell e^{-\alpha\delta}.
\een
The equations (\ref{Delta-liga}) and (\ref{cond-liga}) now allow us to write the important relation
\ben\label{cond-liga2}
\frac{2\Delta}{T_c}=\frac{<\chi>_{eff}}{T_c}.
\een
Recall the examples in Fig.~\ref{fig1} for the effective condensate for three distinct charges in the regime $\alpha\delta\to0$ previously considered. Using the Eq.~(\ref{cond-liga2}) we identify the relations between binding energy  and the critical temperature given by 
$2\Delta\simeq 4\,T_c$, $2\Delta\simeq 8\,T_c$ e $2\Delta\simeq 12\,T_c$.

This seems to point out a behavior of High-$T_c$ superconductors. For the sake of comparison we know that BCS superconductors have a typical relation  $2\Delta\simeq 3.5\,T_c$, whereas the High-$T_c$ superconductors normally enjoy the relation $2\Delta\simeq 5\,T_c$ to $2\Delta\simeq 8\,T_c$.

\begin{figure}[h!]
        \includegraphics[{angle=90,height=6cm,angle=270,width=7cm}]{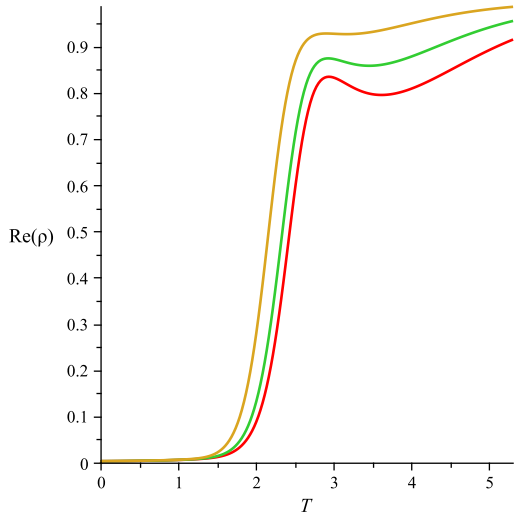}
  \caption{The real part of the AC resistivity at low frequencies as a function of temperature. We use $\delta = 0.40, $ 0.45, and 0.55, from bottom to top; $T_c = 3$, $\omega = 0.8$ and $q = 1$.}
  \label{fig3}
\end{figure}

In the following we present the plot that describes the behavior of the real part of the low frequency AC resistivity $\rho=1/\sigma$ as a function of the temperature --- Fig.~\ref{fig3}. 
Notice that for the resistivity sufficiently above the critical temperature $T_c$ decreases almost linearly with the temperature. Moreover, as the system approaches the critical temperature, the resistivity tends to locally increases, but decreases very quickly below the critical temperature until achieve the resistivity very close to zero. This should be compared with the resistivity versus temperature for three High-$T_c$ superconductor samples of La-Ba-Cu-O with $T_c = 35^ {\rm o}K$ by Bednorz and M\"uller \cite{bm1986}. This result confirms, at least qualitatively, that our superconducting domain walls model agrees with some properties of cuprates which are well-known layer-type perovskite-like structures.
We still note that for values of $\delta$ larger than $0.40$ the system tends to reduce its critical temperature --- see Fig.~\ref{fig3}, the plots for $\delta=0.45$ and $\delta=0.55$. 

Finally, as a last comment we can even improve our above discussions by considering more general orbits.
The original idea of trial orbit method  \cite{bazeia} enables us extending the aforementioned elliptic orbit to general orbits
\ben
\phi^2+{\Big(\frac{\lambda}{\mu}-2\Big)}^{-1}{\chi^n}=a^2.
\een
Although much harder to deal with, this system will produce more general type I and type II solutions in such a way that the condensate may have a more general power 
\ben
<\chi>\simeq T_c\left (1-\frac{T}{T_c}\right)^{1/n}.
\een
This extension may reveal an even more realistic analysis for considering domain wall description of superconductivity. 

\section{Conclusions}
\label{conclu}

In this paper, by considering domain wall description of superconductivity, we have identified a relationship between the binding energy of Cooper pairs and the effective condensate depending on the temperature and the electrical charge.
For charges large enough we get a typical ratio of High-$T_c$ superconductors. We calculate the optical conductivity and show that in the regime of low temperatures and frequencies we get an { infinite DC conductivity}. We conclude that the low frequency AC resistivity as a function of temperature is similar to what happens in High-$T_c$ superconductors. The critical temperature tends to be reduced when we move far from the condensate via the deviation position $\delta$. As future prospects, we intend to attack the problem by investigating other quantities such as the London penetration depth and upper critical field $H_{c2}$ as a function of temperature and effects of anisotropy within the domain walls.


\acknowledgments

We would like to thank CNPq, CAPES, PNPD/PROCAD -
CAPES for partial financial support.

\end{document}